\documentclass[fleqn,12pt]{wlscirep}
\usepackage[utf8]{inputenc}
\usepackage[T1]{fontenc}
\usepackage{bm}
\usepackage[normalem]{ulem}
\usepackage{braket}
\usepackage{mdframed}
\usepackage{color}

\newcommand{\kater}[1]{\textcolor{black}{#1}}

\newcommand{\up}{{\ket{\uparrow}}}
\newcommand{\down}{{\ket{\downarrow}}}
\newcommand{\xket}{{\ket{x}}}
\newcommand{\ketbra}[2]{{\ket{#1}\!\bra{#2}}}

\title{Engineered Dissipation for Quantum Information Science}

\author[1,*]{Patrick M. Harrington}
\author[2]{Erich Mueller}
\author[3]{Kater Murch}
\affil[1]{Research Laboratory of Electronics, Massachusetts Institute of Technology, Cambridge, Massachusetts 02139, USA}
\affil[2]{Laboratory of Atomic And Solid State Physics, Cornell University, Ithaca, NY 14853, USA}
\affil[3]{Department of Physics, Washington University, St. Louis, MO 63130, USA}
\affil[*]{e-mail: pmh@mit.edu}

\begin{abstract}
Quantum information processing relies on precise control of non-classical states in the presence of many uncontrolled environmental degrees of freedom---requiring careful orchestration of how the relevant degrees of freedom interact with that environment. These interactions are often viewed as detrimental, as they dissipate energy and decohere quantum states. Nonetheless, when controlled, 
dissipation is
an essential tool for manipulating quantum information:
Dissipation engineering enables 
quantum
measurement, 
quantum state preparation, and quantum state stabilization. 
The progress of quantum device technology, marked by improvements of characteristic coherence times and extensible architectures for quantum control, has coincided with the development of such dissipation engineering tools which interface quantum and classical degrees of freedom. This Review presents dissipation as a fundamental aspect of the measurement and control of quantum devices and highlights the role of dissipation engineering for quantum error correction and quantum simulation that enables quantum information processing on a practical scale.
\end{abstract}

\begin{document}
\flushbottom
\maketitle
\thispagestyle{empty}

\clearpage
Dissipation makes quantum information science possible. Among other things, it provides the means to measure quantum systems---driving all the paradoxical phenomena that come with entangling 
quantum degrees of freedom with macroscopic states---mapping kets onto cats both dead and alive. When uncontrolled, however, dissipation ruins the sensitive quantum coherences which are at the heart of quantum information science:  Dissipation reduces the fidelity of quantum gates, adds noise to measurement signals, and ultimately poses a challenge to achieving the level of control necessary to harness quantum systems for advantage or insight. Advances in quantum technology must contend with this dual edge---in fact, bend these edges to their favor, to harness---to \emph{engineer} dissipation. In this review, we highlight recent experimental and theoretical advances implementing dissipation, either subtly or bluntly, to advance quantum technologies.

Dissipation engineering principles underlie all quantum 
information processing; any judicious choice of hardware with classical controls must account for naturally accompanying dissipation\cite{call51}. Dissipative system-environment interactions support gate operations and state readout, while in turn, fluctuations of the environment impose quantum coherence limits. \textit{Engineered dissipation}\cite{poya96} incorporates methods that control system-environment interactions, as well as the environment itself, to adapt dissipative processes for tasks including quantum state preparation\cite{bout17,magn18}, stabilization of quantum states \cite{vala06,geer13,holl15,legh15,kimc16,bout17,liu16,lu17,ande19}, entanglement and teleportation of quantum states \cite{PhysRevLett.83.5158, PhysRevLett.91.067901, PhysRevLett.91.097905}, the creation of decoherence-free \cite{bret15} and excitation-number-conserving \cite{haco15, ma19, yana19} subspaces, and the implementation of quantum error detection and correction\cite{ande20,kapi17,legh13,touz18,chen21_google,krin21}. During the present era of noisy intermediate-scale quantum (NISQ) computing\cite{pres18}, practical quantum information processing requires hardware-specific dissipation engineering methods to demonstrate low error-rate devices for scalable quantum computation, simulation, and sensing \cite{altm21}.


In this review we describe the key concepts of dissipation engineering through its applications for quantum information processing and sensing.  In Section~\ref{sec:zeno}, we describe the ``quantum Zeno effect,'' an important paradigm for understanding how dissipation influences quantum dynamics.  In Section~\ref{sec:reservoir}, we detail a broader range of techniques, including state preparation, reservoir engineering, and autonomous feedback.  The remaining sections are focused on applications: quantum error correction (Sec.~\ref{sec:error}), quantum sensing (Sec.~\ref{sec:sensing}), and quantum simulation (Sec.~\ref{sec:simulation}).

\section{Zeno effects and Zeno dynamics}\label{sec:zeno} 

The act of measuring a quantum system can strongly influence its dynamics. In particular, measuring a quantity can prevent it from changing, an effect described as a ``quantum Zeno effect" as an allusion to the ancient Greek paradox \cite{misr77, itan90}. 
The process of repeated measurements introduces measurement \textit{back-action}, a dissipative effect in the system dynamics. In the case of Zeno effects\cite{lane83, kofm00}, back-action dynamics are caused by the measurement process itself, irrespective of particular measurement results \cite{harr17}. Dissipation can be interpreted in terms of ``measurements from the environment'' and Zeno effects are a generic result of dissipation.

The interplay of measurement and quantum dynamics is well-illustrated by a two-level system undergoing Rabi oscillations with frequency $\Omega$ between states $\up$ and $\down$, eigenstates of the $\hat\sigma_z$ operator. \kater{Here we assume that  $\Omega$ is much smaller than the qubit transition frequency as is common in the majority of physical systems.} \kater{If we measure the system repeatedly in this basis after each duration $\tau$}, the system is randomly projected into one of the $\hat{\sigma_z}$ eigenstates. The probability $p$ of flipping from one eigenstate to another between two successive measurements is $p=\sin^2(\Omega \tau)$. Thus the system becomes frozen in one of the $\up$ or $\down$ states when the measurement rate is large compared to the Rabi rate ($\Omega\tau\ll 1$). 
In a system with more coupled levels the implications can be quite rich 
and many dissipative control schemes that leverage measurement back-action can be interpreted through Zeno effects.

\begin{mdframed}[backgroundcolor=black!4]
\vspace{8pt}
\begin{center}
\includegraphics{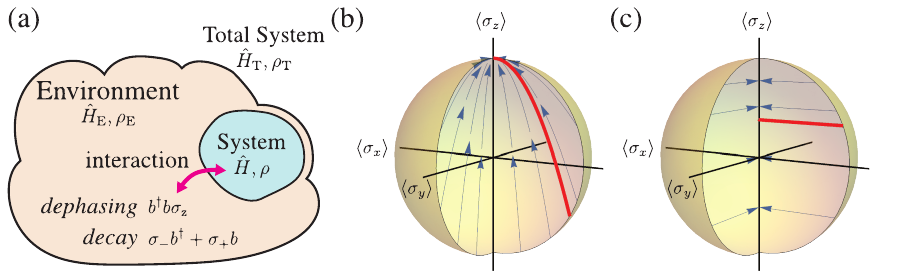}
\end{center}
\vspace{ 8pt}
{\bf Box 1 | Lindblad formalism}

\hspace{8pt} \kater{A typical approach to modeling a dissipative quantum system is to consider it as a subsystem of a larger environment (a). The system bath interaction involves can involve several types of interactions. In the case of decay, flipping the spin $(\sigma_-)$ creates excitations in the bath ($b^\dagger$). In the case of dephasing the energy of the bath modes depend on the state of the system.  The reduced dynamics of the system} are given by equations of motion for the system's density matrix $\hat\rho$, which plays the role of the classical phase space distribution function. Given an ensemble of quantum states $\ket{j}$, which appear with probability $p_j$, the density matrix is $\hat \rho=\sum_j p_j \ketbra{j}{j}$. 
Under several assumptions, (the Born-Markov approximation) where 
the environment can be treated as memoryless and decoupled from the system, the equations of motion for the density matrix has the Lindblad form\cite{breuer},
\begin{equation}\label{eq:lind}
\partial_t \hat\rho=\frac{1}{i}[\hat H,\hat\rho]+\sum_\alpha \Gamma_\alpha \left(
\hat L_\alpha \hat \rho \hat L_\alpha^\dagger -
\frac{1}{2} \hat L_\alpha^\dagger \hat L_\alpha \hat \rho
-\frac{1}{2}\hat \rho \hat L_{\alpha}^\dagger \hat L_{\alpha}\right). 
\end{equation}
The operators $\hat L_\alpha$, referred to as jump operators, describe the dissipation. \kater{As is clear from our formulation, the  coefficients $\Gamma_{\alpha}$ have to be non-negative. This ensures that the system dynamics is a completely positive trace preserving map\cite{breuer}.} In this 
context, dissipation engineering amounts to controlling the jump operators. There can be a rich interplay between the Hamiltonian term and the dissipation, particularly when the the jump operators and Hamiltonian do not commute or the Hamiltonian is time-dependent---a driven-dissipative setting.

\hspace{8pt} \kater{As already described, one} classifies forms of dissipation by reference to the natural basis: For a qubit this is typically the eigenstates of the $\hat \sigma_z$ operator, which commutes with the Hamiltonian of the bare undriven system. In this case, the density matrix is a $2\times 2$ Hermitian matrix, and is fully characterized by the expectation values of the Pauli matrices, allowing it to be visualized as a point on the Bloch sphere.

\hspace{8pt} {\em Decay} involves transitions between basis states. For example, spontaneous emission corresponds to $\hat L\propto \ketbra{\downarrow}{\uparrow}$. As depicted in the Figure panel (b), states evolve to a pole of the Bloch sphere. In contrast, {\em dephasing} does not lead to transitions between basis states, but instead destroys phase-coherence. For example, projective measurement in the energy basis corresponds to two jump operators $\hat L_\uparrow \propto \ketbra{\uparrow}{\uparrow}$, and $\hat L_\downarrow \propto \ketbra{\downarrow}{\downarrow}$. For a qubit, dephasing occurs when $\hat L_\alpha$ commutes with $\hat \sigma_z$. As depicted in panel (c), dephasing brings the state to the $\hat z$ axis of the Bloch sphere without changing $\langle\hat\sigma_z\rangle$.

\hspace{8pt} The term $\hat L_\alpha \hat\rho\hat L_\alpha^\dagger$ corresponds to applying the operator $\hat L_\alpha$ to every state in the ensemble: It encodes the change to the density matrix when a jump occurs. The last two terms in Eq.~(\ref{eq:lind}) represent the influence of the environment on the system in the absence of a jump. They can be combined with the coherent evolution, writing
\begin{equation}
\partial_t \hat \rho =
\frac{1}{i}\left(\hat H_{\rm eff} \hat\rho-\hat\rho \hat H_{\rm eff}^\dagger\right)
+\sum_{\alpha} \hat L_\alpha \hat \rho \hat L_\alpha^\dagger.
\end{equation}
The non-Hermitian effective Hamiltonian,
\begin{equation}\label{eq:heff}
\hat H_{\rm eff} = \hat H-i\sum_\alpha \hat L_\alpha^\dagger \hat L_\alpha,
\end{equation}
represents the evolution of the system conditioned on no jumps occurring \cite{nagh19,el-ganainy,touz18}. The dynamics of this non-Hermitian Hamiltonian describes the reduced dynamics of the system under dissipation, and thereby encodes the Zeno effect (see Sec.~\ref{sec:zeno}). 
\end{mdframed}


The Zeno effect is relevant for engineered dissipation when the coupling to the environment (ie. the measurement rate) is strong.  It leads to the  paradoxical conclusion that increasing the environmental coupling can actually lead to less loss \kater{(of particles, for example)}, in the strong coupling limit.   Consider an incoherent decay from $\down$ to $\up$, corresponding to spontaneous emission of photons with rate $\Gamma$, and described by a jump operator $\hat L=\sqrt{\Gamma} |\!\!\uparrow\rangle\langle \downarrow\!\!|$ (see Box 1). When the dissipation rate is small compared to the Rabi rate, $\Gamma\ll\Omega$, the system rapidly oscillates, and is in state $\down$ approximately half the time. The rate of photon emission is then $\Gamma/2$, which grows with $\Gamma$, as one intuitively expects.  Conversely if $\Gamma\gg\Omega$, the Zeno effect freezes the system in the $\up$ state, and the photon emission rate scales as $\Omega^2/\Gamma$: The dissipation rate actually falls with $\Gamma$. In the limit of strong dissipation ($\Gamma\to\infty$) no photons are emitted and one can replace the dissipation with the constraint that the system can never be in $\down$.  Thus very strong dissipation can be used for coherent control, including the processing of quantum information. 

One way to gain an intuitive understanding of this effect is to interpret it as a complex detuning of the transition from $\down$ to $\up$.
As introduced in Box 1, 
the dynamics conditioned on the absence of an dissipative event are described by 
a non-Hermitian effective Hamiltonian $\hat H_{\rm eff}= \hat H-i \sum_\alpha \hat L_\alpha^\dagger \hat L_\alpha$, where $\hat L_\alpha$ are the jump operators. 
In our example, the term $i\, \hat L^\dagger \hat L = i\,\Gamma \ketbra{\downarrow}{\downarrow}$ gives the excited state a complex energy. The shift of the energy from the real axis yields an effective detuning. If the detuning is large, then any coupling between $\down$ and $\up$ is far off-resonant, and the system stays in the $\up$ state. 

As a simple concrete example, consider a 3-level system, with states $\ket{A}$, $\ket{B}$, $\ket{C}$, and a Hamiltonian which drives transitions $A\leftrightarrow B\leftrightarrow C$ (Fig.~\ref{fig:fig_res_abc}a).  If a very strong dissipation is added to $\ket{C}$, the system will simply undergo Rabi oscillations between $\ket{A}$ and $\ket{B}$, never transitioning to $\ket{C}$ (since the dissipation ``detunes'' that level). For this driven-dissipative system, an effective ground state is formed by the subspace spanned by the states $\ket{A}$ and $\ket{B}$, while excitations out of this ground state are suppressed by the strong dissipation on $\ket{C}$.  The state-selective dissipation introduces a constraint---a feature which is valuable for quantum information processing and sensing.

In this case, the space spanned by $\ket{A}$ and $\ket{B}$ is an example of a {\em dark subspace} or {\em decoherence-free subspace} \cite{PhysRevLett.85.1762}.  Such a space exists whenever there are states in the null-space of the dissipative part  of the effective Hamiltonian, $\hat Q=\sum_\alpha \hat L_\alpha^\dagger \hat L_\alpha$. When the dissipation is strong ($||\hat Q||\gg ||\hat H||$), to leading order, dynamics are restricted to that subspace. Applying the standard prescription for degenerate perturbation theory, the system evolves under $\hat H$ projected into this dark subspace. If $\hat H$ and $\hat Q$ do not commute, the resulting {\em Zeno dynamics} can be rich\cite{facc08,facchi20,lida98,bret15,haco18,sorensen18,mark20,raimond10,raimond12,kumar22,snizko20}.  The remarkable feature here is that strong dissipation can lead to non-trivial coherent evolution. The states in the dark subspace are long lived, with lifetimes that scale inversely with the dissipation strength. 

The Zeno effect gives an explanatory perspective that connects measurement, environmental couplings, and the back-action dynamics from these dissipative processes. Moreover, the Zeno effect offers practical tools for quantum information processing: in atomic systems it is used to extend the life of molecular states \cite{zhu14}, tune interactions \cite{Chen2021}, and even enhance the precision of spectroscopy \cite{ozawa}. Dissipation from strong measurement can create decoherence-free subspaces, which can provide an essential component of quantum error correction \cite{alif08, puri21}. The Zeno effect can even be used for quantum gates, as demonstrated by a recent experiment involving superconducting circuits \cite{Blum21}. There, two qubits had no direct interaction---instead a projective measurement involving an auxiliary state led to entangling Zeno dynamics. 

\section{State Preparation, Reservoir Engineering, and Quantum Measurement}\label{sec:reservoir}
\begin{figure}[h]
\centering
\includegraphics{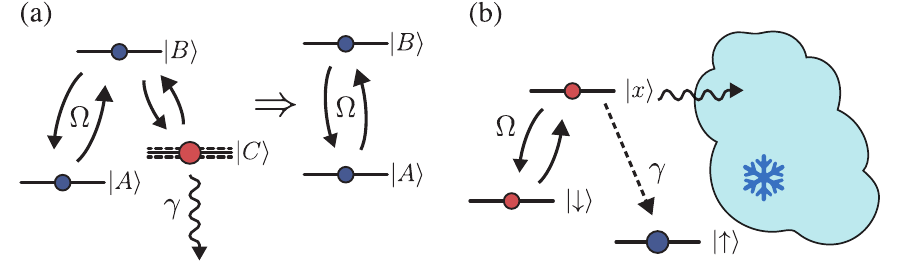}
\caption{(a) The effect of strong dissipation on state $\ket{C}$ creates a decoherence-free subspace spanning the states $\ket{A}$ and $\ket{B}$. (b) An example of optical pumping to a lower energy state $\up$ to achieve ground state cooling. The combination of drive and dissipation irreversibly brings the state $\down$ to $\up$, by coupling to a lossy state $\xket$. Here the environment is a cold reservoir which acts as an ``entropy dump.''}
\label{fig:fig_res_abc}
\end{figure}

While the Zeno effect (Sec.~\ref{sec:zeno}) is largely passive, there are a number of more active approaches to dissipation engineering.
The classic example, illustrated in Figure~\ref{fig:fig_res_abc}b, is optical pumping.  Consider a three-level atom with two long-lived states $\up$, $\down$, and one short-lived state $\xket$.  The latter can decay into $\up$ by emitting a photon. To pump the system into $\up$, one turns on a drive between $\down$ and $\xket$. The system will cycle between those states, eventually decaying into $\up$. If the temperature of the \kater{electromagnetic field environment} is small compared to the level spacing between $\up$ and $\xket$, then the decay is unidirectional; there are no thermal photons to drive the $\up \to \xket$ transition, and hence $\up$ is a dark state.

This simple example highlights a key observation: Dissipation can act as a one-way valve, which inevitably leads the system into a dark-state or manifold. In this  way, entropy is transferred from the system of interest to the bath. This approach has been applied to produce quantum states in all the forerunning platforms for quantum information processing\cite{shan13, kimc16, liu16, lu17, murc12, geer13, magn18, puri17, ande19, mamaev, krauter, souquet, lin, Kienzler53, plenio, krau08,kastoryano,morigi,Ma2021, cole22}.

A wide range of states can be produced by such pumping protocols, including ones that are  highly entangled\cite{aron,rao,kimc16,martin2017,rao2017,schuetz,shao,didier,reiter,ma2019,sharma2021drivendissipative,colladay2021driven,Ticozzi2008}. For example, recent experiments on superconducting qubits and trapped ions\cite{lin,shan13} engineered dissipative processes to prepare a two qubit Bell state, $\ket{\phi_-} \propto \ket{\uparrow \downarrow}-\ket{\downarrow\uparrow}$. This is a maximally entangled state. The requisite engineering was quite involved, taking advantage of detailed features of the hardware, but on a conceptual level it was very similar to optical pumping: Whenever the system is not in the desired target state it is subject to entangling dissipative processes.


A distinct approach to dissipation engineering is illustrated by optical cooling. One can cool motional or low energy internal degrees of freedom by coupling mechanical oscillators\cite{brag70, coha99, chan11, rivi11,chun}, atoms/ions\cite{wine75, hans75,schreck,phillips,stenholm}, or circuits \cite{vala06, murc12} to the electromagnetic field. One needs to arrange a setting where the rate of energy-increasing processes are smaller than those of energy-decreasing processes. 
Such selectivity typically comes from sculpting the density of states, or by adding extra drives which couple to short-lived states.
A classic example is Raman sideband cooling of the motion of a trapped ion \cite{diedrich}.  An optical transition drives the ion from its electronic ground state to a short-lived electronic excited state.  This transition is generically accompanied by  a simultaneous change in the motional wavefunction.  If the drive is red-detuned, then transitions which reduce the motional energy are favored.  This can be interpreted as a form of \textit{coherent feedback}\cite{wise94,lloyd}.

\begin{figure}[h]
\centering
\includegraphics{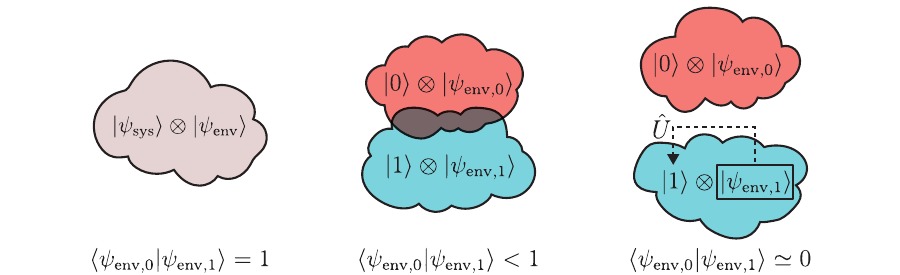}
\caption{\kater{Quantum measurement involves the creation of correlations between a quantum system and an auxiliary ``environment'', indicated here as cloud shapes. The strength of the measurement is dictated by the overlap of the these environment ``pointer'' states. 
The middle figure corresponds to a {\em weak} measurement, where the environmental outcomes are weakly correlated with the quantum state.  In addition to illustrating a {\em strong} measurement, the figure on the right illustrates
feedback, where the evolution of the system (quantified by the operator $\hat U$) is contingent on a certain measurement result.
}}
\label{fig:meas}
\end{figure}

\kater{In all of these cases, aspects of the quantum state are correlated with the environment, which connects closely with modern descriptions of the process of quantum measurement. Here, measurement is treated in a multi-step process. The interaction between the quantum system and its environment leads to changes in the environment that depend on the system's states, as sketched in Figure \ref{fig:meas}.  Here, the environment forms ``pointer states" \cite{zurekpointer}. We use the concept of environment very generally in this case; for example, in the paradigmatic example of a Stern-Gerlach measurement, the interaction couples the electron spin states with its momentum---the momentum degree of freedom plays the role of the environment. After the interaction, a measurement of the pointer state gives information about the system state, collapsing the entanglement. At this point the measurement results become classical information, and the effect of the interaction on the system is referred to as \emph{backaction}. This extends the treatment of measurement beyond textbook concepts of projective measurements. Here the environment can have a much larger Hilbert space than the system (as is the case with the infinite dimensional Hilbert space of the electron's momentum).  As such, the wide range of measurement outcomes from the environment yield different \emph{partial} measurements on the system. These are weak measurements \cite{Jacobs2006}.  The degree of measurement can vary: if the environmental states are partially overlapping (as in the central panel) one only gains partial information about the quantum state \cite{clerk, qnd}.  The right-most panel illustrates the case of a strong, or projective, measurement, where the environmental states are orthogonal \cite{wheeler}; any possible measurement outcome of the environment corresponds to one or the other of the systems states.  The Lindblad formalism in Box 1  ignores the state of the environment, but, as described in Box 2, there are other techniques which allow one to model the measurement outcomes and calculate the system dynamics contingent on those outcomes \cite{Jacobs2006,Hatridge2013,Weber2014,Campagne2016, haco16, flur20}.}

\kater{Quantum trajectories refer to the dynamics of the system over the course of several repetitions of this measurement interaction. Since the Lindblad formalism in Box 1 ignores the state of the environment, the master equation evolution is the result averaging over the ensemble of individual trajectories; correspondingly the quantum trajectories (Box 2) are referred to as the ``unraveling'' of the Lindblad master equation. }

\kater{Quantum measurement embodies the interplay between classical and quantum information. The integration of this classical information into a feedback circuit can alter dynamics\cite{PhysRevA.65.010101}, stabilizing target states of the system \cite{Herasymenko2021}, or optimizing information that is extracted about the system \cite{wiseman, liu16, martin2020}. As we explore in the next section, the interplay of measurement and feedback is critical to the implementation of quantum error correction.}

\section{Quantum error correction}\label{sec:error}

The most important application of the \kater{quantum measurement and feedback} is protecting quantum information. Typical approaches to such quantum error correction utilize {\em stabilizer} or {\em syndrome} measurements---measurements that do not disturb the logical states but provide information which can be used to detect and correct errors. Quantum information is stored in a redundant manner and measurements detect errors at a stage where they can still be corrected. Errors are corrected by gate operations, rotating the system back into the logical computational subspace, or the syndrome measurements are recorded and an appropriate correction is applied at the end of the computation.   The measurement correction cycle is an example of engineered dissipation, precisely because the measurements (which are intrinsically dissipative) need to be engineered not to disturb logical states.  The measurement-feedback system can be implemented at a hardware level, making the system self-correcting \cite{gertler,ahn2002,atalaya2021,kerckhoff2010,kapi16,reiter2017,albe19,sarovar2005,deneeve2020error,kristensen}, or other hardware choices can be made to reduce noise sensitivity \cite{lihm18, gyenis2021}.   Some of these approaches are described as {\em autonomous error correction}, and involve engineered dissipation.

An illustrative example of quantum error correction is the bit flip code\cite{pere85}, which embeds the logical states $\ket{0_L}$ and $\ket{1_L}$ redundantly in three qubits, $\ket{0_L} = \ket{000}$ and $\ket{1_L} = \ket{111}$. A single bit flip error can then be detected by majority vote without revealing the individual qubit states. This is achieved with pair-wise parity measurements\cite{ande19,bult20}. These parity measurements are the prototypical examples of stabilizers---they reveal the occurrence of individual errors, yet, because they commute with all observables of the logical qubits, do not disturb the encoded information. In the language of Section \ref{sec:zeno}, the logical computational subspace is a \emph{dark subspace} in the measurement process.  
Extensions of this simple approach, utilizing a larger quantity of qubits and more complicated stabilizer measurements, in principle allow for arbitrary qubit errors to be corrected.  Different approaches provide varying degree of error tolerance, at the cost of physical resources. For example,  detection of either bit flip or phase flip errors with repetition codes has demonstrated the suppression of logical error rates and favorable scaling on 21 qubits with 50 rounds of quantum error detection \cite{chen21_google}. The surface code has been proposed as a practical approach to large-scale quantum computation\cite{fowl12}, tolerating single qubit error rates comparable to current experimental limits\cite{corc15, ande20} and scalable with current qubit architectures \cite{vers17}.  With present error rates, however, the surface code would still require thousands of physical qubits per logical qubit, leaving the realization of a fault tolerant error corrected quantum processor still a distant experimental goal.

Rather than encode quantum information redundantly in multiple qubits, an alternative approach is to utilize the infinite dimensional Hilbert space of a harmonic oscillator, typically a single mode of a microwave cavity, to realize logical qubits\cite{cai}. The simplest of these {\em bosonic codes} are binomial codes which encode qubits in a finite number of Fock states, $\{\ket{n}\}$, each of which a fixed number of quanta \cite{michael16, axline18}.  The coefficients of the Fock states are related to binomial coefficients, with the minimal example having $\ket{0_L} \propto \ket{0}+ \ket{4}$, $\ket{1_L} = \ket{2}$ as logical qubits. This encoding is chosen so that every logical state has the same parity, such that the loss of a photon, which is the dominant error process for oscillator states, can be detected by measurement of parity. A unitary operation can correct the error without scrambling the quantum information.

Similarly, bosonic Cat codes \cite{legh13,mirr14, albe19, guil19,grimm,gertler,toth} are robust to single photon loss by encoding logical qubits in Schr\"odinger cat states---superpositions of two or more coherent states. Using this architecture, real-time measurement and feedback has demonstrated a logical qubit lifetime longer than the relaxation time of its constituent parts \cite{ofek16}. As a coherent state of an oscillator can be maintained in steady state via a combination of dissipation and resonant driving, the cat states can be stabilized by appropriate two-photon driving and dissipation \cite{legh15, hu19}, in this way, dissipation can be harnessed to actually reduce error rates, beyond just detecting errors for correction.  Extending the approach taken in such cat codes, the Gottesman-Kitaev-Preskill (GKP) error correcting code \cite{gott01} encodes logical qubits in a periodic grid in the phase space of a harmonic oscillator. The GKP encoding is non-local for all three Pauli operators, meaning that small perturbations, entering as small phase space displacements of the oscillator, can be corrected.  In circuit QED, the GKP code has been implemented by creating oscillator displacements conditional on a qubit state in such a way that measurement of the qubit projects the oscillator onto the desired grid state \cite{camp20}.  The GKP state has also been produced using the motional degrees of freedom of trapped ions \cite{fluhmann,deneeve2020error}. 


These examples can all be viewed as digital approaches to error correction, where the evolution is broken into discrete blocks interrupted by stabilizer measurements and feedback. Complementary to this approach is the use of continuous measurement to detect errors \cite{livingston21}, as highlighted in a recent experiment where errors, which take the form of quantum jumps out of a desired space, are detected and corrected through continuous measurement and feedback \cite{Minev2019}. In this experiment, a circuit supporting three quantum levels has one pair of levels that are ``bright" and one state that is ``dark''.  Quantum jumps can take the system out of the bright manifold of states, but continuous monitoring can detect the jumps, enabling a feedback correction to reverse the quantum jump before it is complete.  This feedback correction works because the quantum jumps, while occurring stochastically, correspond to a measurement driven evolution that is coherent. This is because the measurement signals associated with these jumps---darkness---are uniform rather than stochastic. Thus, a feedback controller, detecting only a few moments of dark signal, can apply a rotation to return the three-level system to the bright manifold; the error is corrected even before it has a chance to completely occur.   

Error correction fights fire with fire:  It uses one form of dissipation (measurement) to control unwanted forms of dissipation. When judiciously chosen, the additional dissipation does not disturb the encoded information, but the information gleaned allows a classical controller to compensate for the uncontrolled dissipation.

\hspace{-.3in}
\begin{mdframed}[backgroundcolor=black!4]
{\bf Box 2 | Quantum Trajectories}

\hspace{8pt} The Lindblad master equation in Box 1 models the evolution of the system's density matrix when the state of the environment is ignored:  The state of the environment is traced over.  One may, however, wish to describe dynamics which depend upon the state of the environment.  As a concrete example, consider an atom that is in a superposition of its ground and excited states. If one detects a photon emerging from the atom between time $t$ and $t+\delta t$, then  the density matrix evolves as 
\begin{equation}\label{k1}
\hat\rho_\mathrm{t+\delta t}^\mathrm{click} = \frac{\hat K_\mathrm{click} \rho_t \hat K_\mathrm{click}^\dagger}{\mathrm{Tr}[\hat K_\mathrm{click} \hat\rho_t \hat K_\mathrm{click}^\dagger]},
\end{equation}
where $\hat K_\mathrm{click} = \sqrt{\gamma\,\delta t} \ket{\downarrow}\bra{\uparrow}$ is the {\em Kraus} operator which corresponds to the emission of a photon\cite{kraus1983states}.   Conversely, if no photon is detected the state evolves via the analog of Eq.~(\ref{k1}) but with the operator
$\hat K_{\mathrm{no}\text{-}\mathrm{click}} = \ket{\uparrow}\bra{\uparrow} + \sqrt{1-\gamma\,\delta t}\ket{\downarrow}\bra{\downarrow} $.  In general there are many possible outcomes, indexed by $m$, with Kraus operators $\hat K_m$.   The probability of outcome $m$ is $P(m)=Tr[\hat K_m\rho \hat K_m^\dagger]$, and $\sum_m \hat K_m^\dagger \hat K_m=1$.  
If one averages over all possibilities, one recovers a Lindblad equation with jump operators $\hat L_m= \hat K_m/\sqrt{\delta t}$.
By following individual trajectories one can model both the evolution of a quantum device, and the entire history of the interactions with the environment, or model the quantum dynamics conditioned on certain measurement results.   Such considerations are essential for filtering, post-selection, and real-time quantum feedback.

\end{mdframed}

\begin{figure}[h]
\centering
\includegraphics{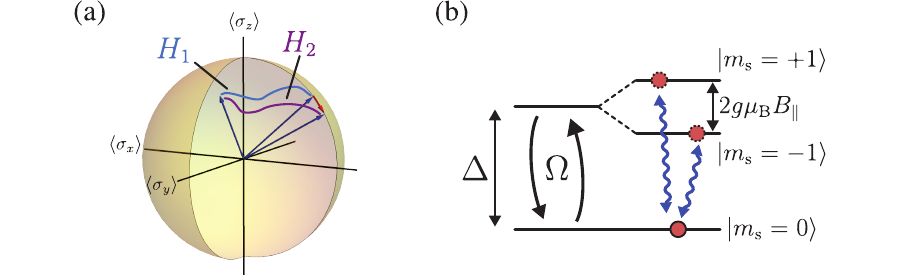}
\caption{(a) In quantum sensing, one determines the parameters that distinguish two different Hamiltonians $H_1$ and $H_2$ through distance between the final states after some time evolution. The sensitivity to the parameters is quantified by the the quantum Fisher information\cite{Braunstein1994,Wootters1981}. (b) Nitrogen vacancy color centers in diamond have optically initializable and readable magnetic sublevels that are particularly suited to sensing. Magnetic field noise induces dissipative transitions between sublevels, making the color center a sensitive spectrometer.  The zero-field splitting $\Delta$ is intrinsic to the material, while the splitting between the $m_s=\pm1$ states is controlled by the magnetic field. Transitions between the states can be driven by a coherent drive ($\Omega$) or magnetic field noise (blue squiggly arrows).}
\label{fig:fig_sensing}
\end{figure}

\section{Application: Quantum Sensing}\label{sec:sensing} 
Quantum mechanics can offer advantages over classical measurement approaches for sensing. First, quantum systems are small.  This gives access to smaller length-scales and boosts sensitivity: the individual energy levels can be sensitive to very weak perturbations. Second, the coherent evolution of a quantum system means that it accumulates phase in proportion to a perturbation of interest, leading to higher precision. Finally, quantum entangled states offer opportunities for reduced noise and enhanced sensitivity \cite{dege17}. 

Typically, one characterizes the performance of a quantum sensor in terms of the quantum Fisher information that can be obtained about a parameter---loosely speaking---quantifying how the distance between two quantum states depends on the sensing parameter of interest (Fig.~\ref{fig:fig_sensing}a).  In this sense, dissipation, which tends to create mixed states and therefore reduces the distance between states, hinders the performance of quantum sensors. Indeed, many protocols for sensing utilize specific (dynamical decoupling) pulses to reduce dissipative effects, while enhancing the desired accumulated phase. Alternatively, error correction approaches can be used to maintain coherent evolution, as is also required in quantum processors. There are, however, some cases where dissipation is an essential aspect of quantum sensing.

A key approach, often employed in quantum sensing using nitrogen vacancy color centers in diamond \cite{maze08, bala08}, is to engineer a situation where the signal is encoded in the dissipation rate. As shown in Fig.~\ref{fig:fig_sensing}b, these color centers have spin sublevels (labeled with quantum numbers $m_s$) with splittings denoted by $\Delta$ and $2g\mu_\mathrm{B} B_\parallel$:  $\Delta$ is the zero field splitting, and $B_\parallel$ is the component of the magnetic field along the quantization axis.  The transitions from $m_s=0$ to the other sublevels are in the microwave frequency range; magnetic field noise resonant with these transitions induces quantum jumps between them. The dissipation rate can therefore be used as a sensitive probe of such magnetic field noise. Recent work utilized this principle to measure the Johnson noise at nanometer distances in normal metal films \cite{Kolk2015}. 

Another application where the spin dissipation can be used as a sensitive probe is in the study of many-body spin dynamics. A recent experiment examined how the polarization of a nitrogen vacancy center can diffuse through interactions with neighboring spins  \cite{zu2021}. The resulting power law decay of polarization gave information which was not accessible through classical probes.

The most common mode of operation for a quantum sensor involves initializing the device, allowing it to evolve, and then measuring. An alternative approach is to continuously probe the quantum system leading to a trade off between the continuous accumulation of information and quantum coherent evolution. When the balance of these two effects is carefully engineered, the resulting driven-dissipative evolution can yield a powerful sensor as detailed below.

A clear example of this balance is demonstrated by dispersive detection of number parity of single electron charges that have crossed the Josephson junction---a sensor that is well-suited to detect the very quasiparticle dynamics that induce relaxation in quantum processors\cite{sern19}. Dispersive readout of charge-parity relies on the dissipative process of single-shot measurement to detect the system's occupation of energy eigenstates with even or odd charge-parity. 
The measurement can be \textit{quantum non-demolition} \cite{Braginsky1980, lupa07}: The system Hamiltonian commutes with the measurement jump operators which cause back-action dynamics. Consequently, dispersive readout ensures that the dissipation does not deleteriously alter the mapping between charge-parity and the readout signal throughout the measurement process. 
As such, the continuous monitoring of charge-parity allows the monitoring other quasiparticle-induced loss that can limit the coherence of superconducting qubits.

As another example, driven-dissipative evolution can be used for low frequency magnetic field detection\cite{xie20}. Here, using again nitrogen vacancy centers, the optical illumination that initiates and reads out the magnetic state of the color center is always on, creating continuous dissipation to the $\ket{m_\mathrm{s}=0}$ sublevel. The combination of this dissipation and an additional microwave drive that couples the magnetic sublevels yields a fluorescence intensity that is proportional to the signal of interest.  The advantage here, is that the sensitivity can be pushed to very low frequencies---circumventing limits posed by the intrinsic coherence of the color center.

This approach of continuous measurement while sensing can also be applied at the level of single quantum trajectories for a single quantum system \cite{flur20, khan21, nola21}.  In this case, a continuous measurement signal can be used to track a quantum system while also gaining information about the parameters of the system's Hamiltonian. This instance highlights the difference between quantum measurement---pertaining to quantum observables---and quantum sensing which pertains to estimating parameters of a system's Hamiltonian.

\section{Application: Quantum simulation}\label{sec:simulation}
Quantum simulation \cite{Blatt,aspuru,gross,houcksim,schafer,noh} refers to using one quantum system to emulate the physics of another \cite{poplavskii75, mani80, feyn82, feyn85}: Neutral atoms hopping on optical lattices stand in for electrons in materials,   superconducting circuits play the role of optical cavities, or atomic Rydberg excitations imitate spins \cite{rydberg,bern17,barreiro}. These platforms introduce new ways of interrogating physics and phenomena that occur on   unaccessible length-scales or timescales.
The same tools that are used for emulation can also be used to create new systems which have no realization in nature.  For example, experiments on superconducting circuits have simulated the behavior of electrons in a hyperbolic geometry \cite{koll19}. A particular interest right now is exploring strongly-coupled models which are not readily analyzed using conventional computational tools \cite{hoen14, sieberer16, foss17}.  
Quantum simulation has myriad near-term applications for in physics, engineering, chemistry \cite{qchem,troyerchem,csim}, and biology \cite{qbio}, for which quantum devices can be tailor-made to emulate a problem of interest.  Engineered dissipation offers convenient methods to control many degrees of freedom and can be an important resource for quantum simulations \cite{ribeiro}. 

\kater{
As should be clear, the scope of
quantum simulation goes beyond faithful replication:  The analog system may be engineered to have all properties of the original system, or one may aim for a subset of those properties. For example, 
dissipative approaches to replicating the ground state of a Hamiltonian may fail to capture the excitation spectrum or more general dynamical aspects
\cite{Wang2017,Sharma2021,PhysRevA.104.032418}.
Such narrowing of the scope of a simulator can help disentangle complicated phenomena, or improve the robustness of its operation.   Furthermore, these simulators can become objects of study, independent of the original motivation.
}

While quantum simulation can  involve ``digital" approaches where  the simulation is performed using gates on a quantum computer, engineered dissipation is most relevant for analog (or hybrid) approaches, where the degrees of freedom of the system-of-interest can map directly onto those of the physical hardware \cite{kjae20}. For analog quantum simulation, dissipation engineering has  several uses:  (1) Dissipation can be used to constrain Hilbert spaces.  Such constraints are particularly important if the analog system has different degrees of freedom than the system-of-interest.  (2)  Dissipation can be used to funnel quantum simulators into states of interest \cite{Sthitadhi20}. The most familiar example of this is cooling, but there also exist protocols in which the dissipation is engineered to pump the system into a particular excited state \cite{ramos2014}. These same tools are also useful for annealing-based computational strategies \cite{anneal}. (3) Engineered dissipation is necessary for studying explicitly dissipative phenomena, such as transport in many-body systems. Here we review the state of the art in each of these areas. Importantly, quantum simulation of many-body systems with dissipative phenomena is largely unexplored in both theory and experiment and holds promise to explore new physics of condensed matter systems\cite{sieberer16, foss17}.

\subsection*{Constraining Degrees of Freedom with Dissipation}
Section~\ref{sec:zeno}  discussed Zeno effects, where measuring a quantity prevents it from changing.  This idea, often supplemented with some sort of coherent feedback (cf. Sec.~\ref{sec:reservoir}), can be used to impose constrains to realize an effective Hamiltonian for quantum simulation.  This parallels approaches to quantum computation where projective syndrome measurements and corrective gates are used to constrain the computer to a chosen code-space (see Sec.~\ref{sec:error}).

\begin{figure}[h]
\centering
\includegraphics{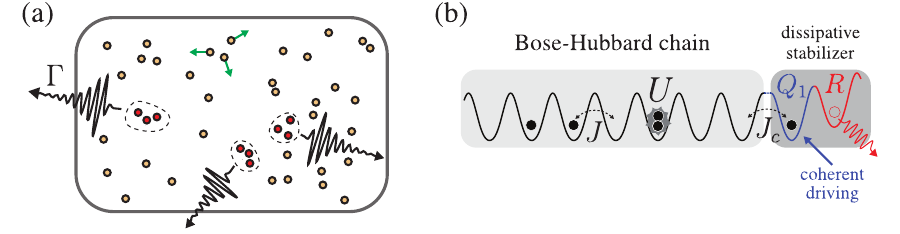}
\caption{(a) An ensemble of molecules have an effective three-body repulsion (green arrows) as a consequence of strong three-body loss (black arrows). (b) The dissipative stabilization of a Mott insulator state, with one particle per site. Particle number is a conserved quantity in the effective ground state due to energy selective dissipation and the incompressibility of the Mott insulator state. \textit{Figure adapted from Ref.~\cite{ma19}.}}
\label{fig:fig_sim}
\end{figure}

To illustrate the usefulness of using dissipation to implement constraints, consider ongoing
attempts to produce cold atom analogs of the fractional quantum hall \cite{fqh} state known as the ``Pfaffian" state  \cite{pfaffian}.  This is a topologically ordered state, first discussed in the context of the fractional quantum Hall effect of 2D electrons in large magnetic fields \cite{fqh}. It supports Majorana fermion excitations, and is the exact ground state of a model Hamiltonian with extremely strong short-range three-body interactions. Thus a key step in producing this state is to engineer a strong three-body repulsion (Fig.~\ref{fig:fig_sim}a). Producing such many-particle interactions is quite challenging.  Nonetheless, it is straightforward to create a strong three-body loss term \cite{roncaglia,daley3}, for example by tuning near a scattering resonance. As emphasized in Sec.~\ref{sec:zeno}, if this three-body loss is strong enough, the Zeno effect will restrict the system to the desired manifold, where three particles never come near each other (see Fig.~\ref{fig:fig_sim}). In the presence of an appropriately tuned gauge field, the ground state with this constraint is the desired Pfaffian.
Beyond this example of  state preparation, the behavior of systems with strong three-body losses can be quite rich \cite{diehl3,diehl3b,dogra,zundel,bonnes}. 

More generally, one adds a constraint by inducing large loss:  Large two-body loss induces a strong effective two-body interactions \cite{syassen,froml,tomita2,durr}; Large three-body loss induces a strong effective three-body interactions.
Variants of this basic motif  have been explored in contexts ranging from implementing magnetic models \cite{chiral} to simulating gauge theories \cite{banuls, stannigel2014,caballar}.   The more sophisticated versions of this strategy involve engineering the dissipation so that it actively pumps the system into the constrained space:  This can be through an autonomous feedback scheme (Sec.~\ref{sec:reservoir}) or an active approach involving measurements and correction (Sec.~\ref{sec:error}). The states which satisfy the constraint become part of a dark state manifold.  

An important application is using dissipation to constrain particle number.  This allows one to emulate the behavior of conserved particle-number, such as atoms or electrons, with entities whose number are not conserved, such as photons or phonons. 
For concreteness, the behavior of atoms moving around on an optical lattice can be emulated by the photonic excitations of superconducting circuits \cite{houcksim, caru20}. In that context, one wants to find a dissipation mechanism which ``measures'' the number of photons such that the excitation number is stabilized: injecting more if the number is below the target, or removing excitations if the number is too large.  In the language of statistical mechanics, one can think of this as creating an environment with a finite chemical potential for photonic excitations. 

A number of autonomous schemes have been proposed to produce an effective chemical potential for photonic excitation. The most direct approach has been implemented in experiments where the excitations of dye molecules are used as photon bath \cite{dye}. There are also strategies involving parametrically oscillating the coupling between a photonic system and its bath \cite{hafe15}.  One of the most important insights is that it often suffices to apply the stabilization locally at only a single discrete location:  As long as the excitations are sufficiently mobile, fixing the density locally will fix the average atom number.

This insight is illustrated by an experiment that reports the  autonomous stabilization of a ``Mott insulator'' in a superconducting circuit (illustrated in Figure~\ref{fig:fig_sim}b) consisting of eight coupled anharmonic quantum oscillators (transmons) coupled to microwave resonators \cite{ma19}.  The number of quanta on each device is analogous to the number of particles on a site---and the goal is to have exactly one particle on each site (which is the defining feature of an ideal Mott insulator).  The site at one end, denoted $Q_1$, is coupled to a ``cold reservoir'' realized by a lossy resonator denoted as $R$. The end-site $Q_1$ is driven such that it is forced into a configuration with exactly one excitation. Given the ability of the excitations to hop between sites, the configuration with one particle per site is a dark state.  If there is a particle ``hole'', or lack of an on-site excitation, then excitations will propagate until the hole travels to $Q_1$, where it will be removed.
The advantage of introducing local dissipation at a single site is that it is both easier to implement and leaves an unperturbed ``bulk.'' 
However, the disadvantage is that the time it takes to remove a hole grows with the system size---excitations out of the engineered ground state may not be strongly suppressed.

This example of single-site dissipation highlights the importance of spatial structure. In many cases the most easily implemented dissipation elements are local. This introduces some constraints on the type of states one can produce. There are many examples in the literature of ideas for producing matrix product states or pair entangled states \cite{vers09, krau08,Jaschke_2018, yanaychiral}, including condensates, $\eta$-condensates, pair condensates, and dimerized phases \cite{diehl,pwave,cian19,ramos2014}.  Due to their topical nature, particular efforts have been made to come up with approaches to produce states which either exhibit topological order, or have topologically nontrivial band-structure \cite{bardyn,shavit,pwave,goldstein,iemini,ozawa,dangel,barbarino}. Despite the dissipation being local, these systems exhibit globally conserved quantities. 
There are analogies between these nonlocal degrees of freedom and the protected logical qubits of quantum error correcting codes \cite{kitaev}.



In principle, any thermodynamic quantity can  be constrained by using similar techniques to those of the Mott insulator experiment.  Here, dissipation can introduce an effective chemical potential and analogous approaches would correspond to the appropriate conjugate variable:  For example, constraining the total spin of a system would introduce an effective magnetic field.  


\subsubsection*{Simulating Dissipative Systems}
In addition to being a tool for implementing constraints and projecting into desired states, engineered dissipation can be used to emulate and study exotic dynamics of open quantum dynamical systems.  
One important class of such studies is the imitation of thermal baths or reservoirs. Thermal ensembles have obvious physical importance and they  have use in numerical algorithms such as optimization \cite{pincus} and machine learning \cite{bishop}. A straightforward way to simulate a thermal system is to directly implement a large reservoir with many degrees of freedom \cite{demarco,lena,mckay}.  This is resource intensive, which has motivated approaches where a small number of driven lossy degrees of freedom leads to a thermal ensemble \cite{shabani, metcalf, su, schonleber, dive, griessner,yanayres}.  The governing principle in engineering these artificial thermal baths is the same as used for numerical calculations: A steady-state Boltzmann distribution will be found if the detailed balance condition is satisfied, the rate for transitioning from state $i$ to state $j$, $P_{i\to j}$, is related to the reverse rate by the energies of the two states: $P_{i\to j}/P_{j\to i}=e^{\beta (E_i-E_j)}$, where $1/\beta=k_B T$. \kater{We emphasize that this condition must be engineered, and a generic dissipative system will not satisfy detailed balance.}
Examples which use this principle include coupling superconducting qubits to lossy driven microwave resonators \cite{shabani}, or driven lossy qubits \cite{metcalf}. Traditional optical cooling techniques can be considered as special cases \cite{Liu_2013,Letokhov_1995,marquardt,McKay_2011,guo}. Note that the resulting steady-state properties from these approaches will be universal, but the way the system approaches equilibrium will depend on the details of the reservoir and couplings. There are, however, strategies for emulating generic Lindblad equations, which can fully model the equilibration process \cite{shen17,candia,chenu,sweke,zanardi}.
The thermal baths engineered with these techniques can have a range of tunable parameters: One can engineer both how they couple to the system,  the spectral density of states, and the extent to which information can be stored in the reservoir
\cite{carmele,lebreuilly17}.  


The most novel studies involve emulating non-thermal open quantum systems---largely with the goal of observing new phenomena. This includes 
a range of exotic {\em non-equilibrium phases} \cite{spinliquid, maghrebi, hurst} and
non-equilibrium analogs of equilibrium phase transitions \cite{tomita,dogra,rylands,sieb13,essink,scarlatella,matheycrit,Brennecke11763,foss-feig,diehl,fitzpatrick}.  \kater{These open quantum systems display a similar richness as classical dynamical systems, including limit cycles, period doubling\cite{else,Sacha_2017,rieracampeny}, and all of the complexity which is found in actively driven\cite{buca} and even living systems\cite{Marais2018}. }
They also show purely quantum phenomena such as collapses and revivals \cite{carmele}. \kater{The richness of this behavior can be used in reservoir quantum computing \cite{qrc1,qrc2}, where dissipation is valuable for its contribution to the fading-memory property\cite{carroll2022}.}



These examples illustrate the value of quantum simulation, where one leverages the controllability of one quantum technology to peer into systems that are more difficult to probe. In this endeavor, dissipation provides a range of techniques to adapt one type of quantum system to the physics contained in a desired Hamiltonian.

\subsection*{Conclusions and perspectives}
In the past decades, progress in quantum technology has been marked by increasing control,  particularly regarding the strength and nature of the coupling to the environment. This has led to fundamental and foundational advances in quantum science. This review has focused on cases where deliberate coupling to the environment yields substantive advantages. Such an approach may appear counter-intuitive at first; one might expect coupling to an environment to increase a system's entropy. Indeed, much of the progress in quantum information processing has been due to reducing coupling to uncontrolled degrees of freedom in the environment \cite{koch07, manu09, gyen21}. Nonetheless, judicious engineering of an environment can resourcefully reduce a system's entropy. There are a number of divergent strategies:  In some cases the environment is effectively very cold, as with the example of optical cooling discussed in Sec. \ref{sec:reservoir}, and hence acts as an entropy dump. In other cases, such as when a system is being continually measured, the environment formally takes the form of an infinite temperature bath. The information gained from the measurements, however, can be used to reduce the entropy. The prime example of this approach is quantum error correction.

Dissipation also provides new mechanisms for coherent control. An overarching strategy is provided by the quantum Zeno effect, where strong dissipation imposes constraints on the system dynamics.  This can be interpreted in terms of detuning the system's eigenenergies on the complex plane, leading to Zeno effects and Zeno dynamics within a protected subspace. Even more control can be achieved with autonomous feedback, where the addition of coherent driving can funnel states into a protected subspace. 


While this review has largely focused on practical issues, newfound capabilities to engineer many-body quantum system systems has motivated further exploration of these fundamental concepts. The first of these is \emph{quantum thermodynamics}, which is an emerging field of physics where concepts in quantum information are united with thermodynamic principles such as entropy, heat, and work  \cite{deffner,Vinjanampathy}. Quantum thermodynamics provides a framework to further understand and engineer dissipation.


Similarly, there is a revolution in \emph{quantum dynamical systems} \cite{polkovnikov}. These differ from their classical counterparts due to the structure of the underlying microscopic equations, but also due to the importance of quantum entanglement \cite{maldacena,nahum,lifisher}. Deep insight is being developed into the connections between classical and quantum chaos \cite{alessio}, how information propagates in a quantum system \cite{luitz}, and the interplay between coherent and incoherent processes in  the propagation of entanglement \cite{chan,lichenfisher}. There are novel dynamical phase transitions with universal critical behavior \cite{skinner,sieb13,ippoliti,marino21}. Finally, at the intersection of quantum dynamical systems and quantum thermodynamics are questions about equilibration, when quantum systems can be described thermodynamically \cite{deutsch,nandkishore,abanin,khemani}, and quantifying the information complexity of such systems \cite{vald17, wals21}. The developments which are enabling quantum computation have not only presented these questions, but offer new tools to understand them experimentally.

\kater{Techniques for modeling the dynamics of open quantum systems are continually evolving.  Frontiers include techniques using tensor networks \cite{Jaschke_2018,zwolak,cui2015} or neural networks \cite{nagy,hartmann,vicentini,yoshioka,liu}.  A difficult challenge is going beyond the Markov and Born approximations which were at the heart of much of our discussion \cite{Vega,Vacchini2019}.  
The bath is not necessarily weakly coupled to the system; it can act as a memory, which is entangled with the system in non-trivial ways.  If mastered, this complexity can be a resource.}

It is clear that engineered dissipation is a key part of the technology of quantum information science.  The importance of these concepts will only grow over the coming years. 


\section*{Acknowledgements}
We thank Max Hays, Peter McMahon, Agustin Di Paulo, and Yariv Yanay for critical comments.
This research was supported by an appointment to the Intelligence Community Postdoctoral Research Fellowship Program at MIT, administered by Oak Ridge Institute for Science and Education through an interagency agreement between the U.S. Department of Energy and the Office of the Director of National Intelligence. This material is based upon work supported by the National Science Foundation under Grant No. PHY-2110250, No. PHY-1752844 (CAREER), and a New Frontier Grant awarded by the College of Arts and Sciences at Cornell. 






\clearpage
\bibliography{references}

\end{document}